# Brain MR Image Segmentation in Small Dataset with Adversarial Defense and Task Reorganization

*Xuhua Ren, Lichi Zhang, Qian Wang\*, Member, IEEE and Dinggang Shen\*, Fellow, IEEE*

**Abstract.** Medical image segmentation is challenging especially in dealing with small dataset of 3D MR images. Encoding the variation of brain anatomical structures from individual subjects cannot be easily achieved, which is further challenged by only a limited number of well labeled subjects for training. In this study, we aim to address the issue of brain MR image segmentation in small dataset. First, concerning the limited number of training images, we adopt adversarial defense to augment the training data and therefore increase the robustness of the network. Second, inspired by the prior knowledge of neural anatomies, we reorganize the segmentation tasks of different regions into several groups in a hierarchical way. Third, the task reorganization extends to the semantic level, as we incorporate an additional object-level classification task to contribute high-order visual features toward the pixel-level segmentation task. In experiments we validate our method by segmenting gray matter, white matter, and several major regions on a challenge dataset. The proposed method with only seven subjects for training can achieve 84.46% of Dice score in the onsite test set.

## 1 Introduction

Brain MR segmentation plays a pivotally important role in clinical diagnosis. But manual segmentation is generally time-consuming and prone to errors due to inter- or intra-operator variability. Therefore, fully automated segmentation method is essential for brain image analysis, which has also drawn a lot of attention from the community. The MRBrainS18[1] challenge, for example, aims to find the optimal algorithm for automatic segmentation of individual regions of interest (ROIs) in 3T brain MR scans, including gray matter, basal ganglia, white matter, white matter lesions, cerebrospinal fluid, ventricles, cerebellum and brain stem. The challenge reflects a common circumstance, as the number of the labeled images available to training is often limited. Also, the ROIs in the brain can vary a lot in position, appearance, scale, and etc., making it hard for a single network to adapt to all segmentation tasks.

With the success of deep learning in medical imaging, supervised segmentation approaches built on 3D convolutional neural networks (CNNs) have produced superior segmentation result with satisfactory speed performance. Dolz et al. proposed Hyper-DenseNet [1], a 3D fully convolutional network (FCN), and extended the definition of dense connectivity to multi-modal brain segmentation. Roulet et al. [2] focused on training a single CNN for brain and lesion segmentation using heterogeneous datasets. Wang et al. [3] investigated how test-time augmentation can improve CNN's performance for brain tumor segmentation and used different underpinning network structures.

[1] https://mrbrains18.isi.uu.nl/

Although these studies provide new paradigms in automatic brain segmentation, it is still difficult to handle the large variation of the neuroanatomical structures from different individuals using only limited number of training subjects. Moreover, the task-level interaction with the prior knowledge of brain anatomies is mostly ignored. To this end, we propose a novel CNN based segmentation method for brain MR image segmentation in this work. Specifically, we deploy the adversarial defense framework, such that the adversarial examples combined with the original small training dataset can improve the robustness of the network. Then, we split the to-be-segmented brain ROIs into several groups according to the anatomical prior knowledge, and thus reorganize the segmentation tasks. Whereas the pixel-level segmentation task further benefits from an additional object-level classification task, which helps ease the difficulty in segmentation.

We summarize our main contributions as follows:

1) We deploy the adversarial attack and defense framework on medical image segmentation, to alleviate the concern of small training dataset.

2) We propose a task reorganization strategy to group the brain ROIs according to anatomical prior knowledge.

3) We extend task reorganization to different semantic levels by incorporating an additional object-level classification task, such that the classification and segmentation tasks can interact with each other toward better segmentation performance.

## 2 Methods

In this section, we detail the proposed method for automatic brain MR segmentation, which is featured by two innovative components: **Adversarial Defense Module** and **Task Reorganization Module**. The whole pipeline of our method is shown in **Fig. 1**. The adversarial defense module is designed for generating adversarial examples by Fast Gradient Sign Method (FGSM) [4], aiming at increasing the robustness of the trained model (Section 2.1). For the task reorganization module, we first split the segmentation tasks of 8 ROIs (**Fig.1**) into three groups based on the anatomical prior knowledge, such that all segmentation tasks can be solved from easy to hard (Section 2.2). Moreover, we incorporate an additional object-level classification task, which determines whether a certain ROI exists given a specific 2D slice. The object-level classification task is properly organized within the segmentation network, such that the pixel-level segmentation task benefits from the high-level semantic information learned in classification. Details of the task reorganization module are presented in Section 2.2.

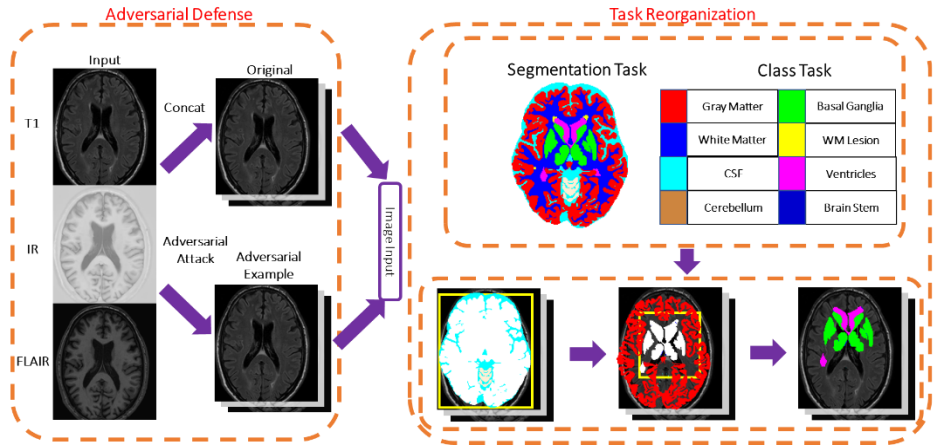

**Fig. 1.** Illustration of our proposed method. 1) Adversarial defense which combines adversarial examples with the original images to increase the robustness of model; 2) Task reorganization which reorganizes the multi-class segmentation task into several simple tasks following anatomical information. Moreover, it also reorganizes the single segmentation task into pixel-level segmentation and object-level classification tasks.

### 2.1 Adversarial Defense

Adversarial attack causes a neural network to deviate from its correct prediction. An adversarial sample is often generated by modifying a real input very slightly, in order to fool a machine learning classifier toward misclassification. In many cases, these modifications can be so subtle that a human observer even cannot notice them at all. However, many existing machine learning classifiers are highly vulnerable to the attacks from adversarial samples. Considering an original input $x$, for a neural network $f$ parametrized by $\theta$ which maps the input $x$ to the output $y$, we can define an adversarial perturbation $r$ as follows:

$$r = \arg\min |r|_2 \quad \text{s.t.} f(x + r; \theta) = y_t, \tag{1}$$

where $y_t$ is the target label of the adversarial sample $x_{adv} = x + r$ and thus differs from the label of the original input $x$.

The adversarial defense strategy can be naturally generalized to the case of image segmentation, where the network is trained with independent cross-entropy loss $L$ at each pixel. By regarding the segmentation network as $f$, we apply FGSM to generate adversarial samples by adding the perturbation $r$ into the input $x$ as:

$$x_{adv} = x + r = x + \epsilon \cdot \text{sign}(\nabla_x L(f(x; \theta), y)). \tag{2}$$

This is a single-step attack, which constrains perturbation range by the parameter $\epsilon$.

Kurakin et al. [5] studied the adversarial defense in training, which generates adversarial samples online and adds them into the training set. They have found that training with adversarial samples generated by the single-step method can improve robustness to other attacks, while the performance difference is negligible if tested with clean inputs. That is, the robustness of the network has been improved. In this paper, we mix single-step adversarial samples ($x_{adv}$) with the original inputs for training, which then effectively augments the size of the training dataset and increases the robustness of our segmentation network.

### 2.2 Task Reorganization

**Splitting segmentation tasks from easy to hard.** In the task reorganization module, we first group different segmentation tasks corresponding to the 8 ROIs, which is the objective in the MRBrainS18 challenge. The ROI labels are presented in **Fig.1**, where we design 3 individual classifiers to solve the segmentation from easy to hard. Specifically, we utilize the first classifier to segment the brain stem (*BS*), cerebellum (*C*) and cerebrospinal fluid (*CSF*), all of which occupy large portions in human brain. The first classifier also predicts the *locating* class (white region in **Fig.1**) which merge the gray matter (*M*), white matter (*WM*), white matter lesions (*L*), basal ganglia (*B*), and ventricles (*V*) into one class. The cropped region of the input images based on the bounding box (yellow box in **Fig.1**) which is obtained by the *locating* class is used as the input of the second classifier for *M, WM* and *L* segmentation. After the second classifier finishing the task, image region inside the other tissues, i.e., *B, V* which is obtained by the *locating* class in the second classifier is used as the input of the third network for *B* and *V* class segmentation. Here is our task reorganization module in object-level representation.

**Joint learning of pixel-level segmentation and object-level classification.** Moreover, we aim to exploit the tasks of different semantic levels in the task reorganization module. While state-of-the-art segmentation methods often require large number of training data, it is possible to utilize the mutual dependency of the tasks of multiple semantic levels to ease the challenge faced by FCN for segmentation. Here, we propose two tasks, i.e., (1) pixel-wise image segmentation and (2) object-level classification which predicts the object classes within the image. These two sub-tasks are closely coupled, i.e., object-level classification is the high-level representation of pixel-wise image segmentation task.

For the high-level class tasks, we adopt the classification loss designed as:

$$L_{cla} = \sum_{x \in \Omega} \text{BCE}(C(x), \hat{C}(x)), \tag{3}$$

where BCE() is the binary cross entropy, $C$ and $\hat{C}$ are the ground-truth and estimated class labels for a single image $x \in \Omega$, and $\Omega$ is the training image set.

Moreover, we utilize a hybrid loss $L_{seg}$ to supervise the low-level segmentation task:

$$L_{seg} = \sum_{x \in \Omega} \text{CE}(Y(x), \hat{Y}(x)) + (1 - \text{Dice}(Y(x), \hat{Y}(x))), \quad (4)$$

where $\text{CE}(Y(x), \hat{Y}(x))$ is the cross entropy between the segmentation ground truth $Y$ and the estimated $\hat{Y}$ for a single image $x \in \Omega$, and $\Omega$ is the training image set. $\text{Dice}(Y(x), \hat{Y}(x))$ is the mean Dice score between $Y$ and $\hat{Y}$. The existence of a certain class in the image has strong correlation with the accuracy of pixel-wise segmentation task. So, the object-level classification task would have benefit for the segmentation task.

Both of object-level and semantic-level representations in task reorganization module could ease the difficulty of the main task by reorganizing the original task to several simple sub-tasks.

**Table 1.** Comparisons of baseline with our proposed modules.

|  | Dice: % |
|---|---|
| Evaluation of Adversarial Defense | |
| Base + Class + Defense ($\epsilon$: 0.05) | 82.42 |
| Base + Class + Defense ($\epsilon$: 0.1) | 82.71 |
| Base + Class + Defense ($\epsilon$: 0.2) | 82.31 |
| Evaluation of Task Reorganization | |
| Base | 81.89 |
| Base + Class | 82.44 |
| Base + Class + Defense | 82.71 |
| **Base + Class + Defense + Coarse2Fine (Proposed)** | **83.12** |
| Comparison with State-of-the-Art Methods | |
| Unet [6] | 81.95 |
| Vnet [7] | 81.88 |
| **Base + Class + Defense + Coarse2Fine (Proposed)** | **83.12** |

## 3 Experiments and Results

In this paper, we use MRBrainS18 challenge data to validate our method. The dataset has 21 subjects acquired on a 3T scanner at the UMC Utrecht (the Netherlands). Each subject contains fully annotated multi-sequence (T1-weighted, T1-weighted inversion recovery, and T2-FLAIR) scans. The subjects include patients with diabetes, dementia and Alzheimers, and matched controls (with increased cardiovascular risk) with varying degrees of atrophy and white matter lesions (age > 50). We use 7 subjects as our training and validation sets. Participants in this challenge were required to submit the segmentation results for onsite evaluation. The test data (14 subjects) were not released to the public. We utilized 5-fold cross-validation in hyper-parameters tuning. The network was trained and applied with a Titan X GPU on Tensorflow [8] and NiftyNet [9] platform.

### 3.1  The Role of Adversarial Defense

**Table 1** shows the performance incurred by the adversarial defense module. First, we consider the single task of segmentation only ("*Base*"), derived from Cascaded Anisotropic Convolutional Neural Networks [10]. We also added the proposed semantic-level task reorganization ("*Class*") in this experiment. Then, we find mixing adversarial samples with the original dataset ("*Base + Class + Defense*") makes the network more robust to the clean inputs, compared with *("Base + Class")* model. In particular, we tuned the parameters of $\epsilon$ in FGSM between 0.05 and 0.2. In both cases, the value of 0.1 shows greater performance, which is selected as the default parameter. Therefore, we conclude that the proposed adversarial defense module could improve the performance of model.

### 3.2  The Role of Task Reorganization

To validate the design of our framework with two levels of task reorganization representations, i.e., object-level ("*Coarse2Fine*") and semantic-level ("*Class*"), we compare several different settings and report the results in **Table 1**. First, we consider the single task of segmentation only ("*Base*"). Second, we further verify the contribution of the proposed semantic-level task reorganization module. We add the class tasks ("*Class*"); the experimental results in the middle of **Table 1** show that the semantic-level representation solution outperforms the single-task solution. Third, with the object-level representation ("*Coarse2Fine*") added in the framework, the combination solution to semantic segmentation outperforms the ("*Base + Class + Defense*") model. Therefore, we conclude that the proposed task reorganization module is beneficial to the segmentation task.

### 3.3  Comparison to State-of-the-Art Methods

Finally, we compare our proposed method with other state-of-the-art algorithms (Unet and Vnet) in **Table 1**. The results show that the proposed method outperforms all the methods under comparison in the validation set. We have also provided visual inspection of the typical segmentation results (Unet *vs.* our proposed method) with the ground truth in **Fig. 2**. The labeling result of the region inside the yellow box shows that, with the integration of our proposed module, the labeling accuracy of our model is improved.

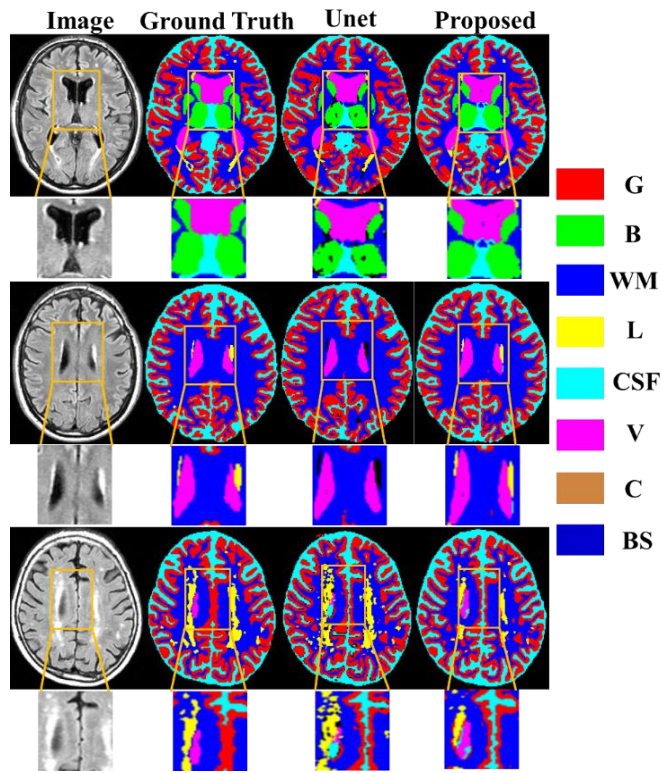

**Fig. 2.** Example segmentation results for MRBrainS18 by using the Unet and our proposed methods. Our proposed method produces significant accurate labels for the regions inside the yellow box.

Furthermore, we compare our method with the winner solution to the MRBrainS18 challenge in **Table 2**. Though our method falls short of the top-two ranked teams, it is worth noting that our submission is indeed corresponding to the "*Base + Class*" model in **Table 2**. In this work, we have further substantially improved our method and achieved much better segmentation results, i.e., comparing "*Base + Class + Defense + Coarse2Fine*" to "*Base + Class*" in **Table 1**.

**Table 2.** Comparisons of the intermediate method result with other top-rank results submitted to on-site validation of MRBrainS18.

| Team | Dice: % | | | | | | | |
|---|---|---|---|---|---|---|---|---|
| | G | B | WM | L | CSF | V | C | BS |
| MISPL | 86.0 | 83.4 | 88.2 | 65.2 | 83.7 | 93.1 | 93.9 | 90.5 |
| K2 | 85.1 | 82.3 | 87.5 | 61.6 | 83.3 | 93.4 | 93.5 | 91.1 |
| "Base + Class" | 85.6 | 78.3 | **89.0** | 60.9 | **83.7** | 92.9 | **94.2** | **91.2** |

## 4    Conclusion

We have proposed an effective adversarial defense, task reorganization framework for semantic segmentation of brain MR image segmentation. Specifically, we proposed adversarial defense to penalize the noise and variance in small dataset for improving the robustness of network. Moreover, we reorganize multi-class segmentation to several sub-tasks following anatomical mechanism. Finally, we reorganize the very challenging semantic segmentation task to several sub-tasks, which are associated with low-level to high-level representations. We have conducted comprehensive experiments on popular medical image datasets. Our results are currently top-ranked in the challenge; specifically, we only addressed intermedia model ("*Base + Class*") in the challenge.